\documentclass[useAMS,usenatbib]{mn2e}
\usepackage{graphics}
\usepackage{epsf}
\usepackage{subfigure}
\usepackage{amsmath}
\usepackage{amssymb}

\def\ni{\noindent}

\begin{document}

\title[Cepheid PC \& AC relations - V]{Period-colour and amplitude-colour relations in classical Cepheid variables V: The Small Magellanic Cloud Cepheid models}
\author[Kanbur et al.]{Shashi M. Kanbur$^{1}$\thanks{E-mail: kanbur@oswego.edu}, Chow-Choong Ngeow$^{2}$ and Greg Feiden$^{1}$
\\
$^{1}$Department of Physics, State University of New York at Oswego, Oswego, NY 13126, USA
\\
$^{2}$Department of Astronomy, University of Illinois, Urbana-Champaign, IL 61801, USA
}

\date{Accepted 2007 month day. Received 2007 month day; in original form 2007 March 15}

\maketitle

\begin{abstract}

Period-colour (PC) and amplitude-colour (AC) relations at maximum, mean and minimum light are constructed from a large grid of
full amplitude hydrodynamic models of Cepheids with a composition appropriate for the SMC (Small Magellanic Cloud). We compare these
theoretical relations with those from observations. The theoretical relations are in general good agreement with their observational
counterparts though there exist some discrepancy for short period ($\log [P] < 1$) Cepheids. We outline a physical mechanism which can, in principle, be one
factor to explain the observed PC/AC relations for the long and short period Cepheids in the Galaxy, LMC and SMC. Our explanation relies on the hydrogen ionization front-photosphere interaction and the way this interaction changes with 
pulsation period, pulsation phase and metallicity. Since the PC relation is connected with the period-luminosity (PL) relation, it is postulated that such a mechanism can also explain the observed properties of the PL relation in these three galaxies.

\end{abstract}


\section{Introduction}

The papers in this series are devoted to a study of the period-colour (PC) and amplitude-colour (AC) relations for classical Cepheid variables, the physics behind the connections of PC and AC relations, and the implication for Cepheid period-luminosity (PL) relations. The connection between the PC and AC relations was derived by \citet{sim93}, who applied the Stefan-Boltzmann law at the optical maximum and minimum light together with the fact that radial variations are small in the optical \citep{cox80}. Specifically:  

     \begin{eqnarray}
       \log T_{max} - \log T_{min} = \frac{1}{10}(V_{min} - V_{max}),
     \end{eqnarray}

\ni where $T_{max}$ and $T_{min}$ are the effective temperature at the maximum and minimum light, respectively. If $T_{max}$, and hence the colour at maximum light, is independent of period, then equation (1) implies there is a relation between the temperature at minimum light and the optical amplitudes, and vice versa. \citet{cod47} found that at the maximum light, the spectral type of Galactic Cepheids is independent of pulsating period. Convincing observational support for a flat (i.e. zero slope) maximum light PC relation has been documented by \citet{sim93}, \citet[][hereafter Paper I]{kan04}, \citet[][hereafter Paper II]{kan04a} and \citet[][hereafter Paper III]{kan06} for the Galactic (with $\log [P]>0.8$, where $P$ is pulsation period in days) and LMC (with $\log [P]>1.0$) Cepheids. These authors imply that the flatness of the Cepheid PC relation at maximum is due to the interaction between the hydrogen ionization front (HIF) and the photosphere \citep[defined as optical depth ${\tau}=2/3$; see Paper II, III and][]{sim93}. However, Paper I found that the SMC PC relation, in $(V-I)$ colour, at maximum light is not flat, but instead suffers
a "flattening" of the slope from $0.396\pm0.039$ to $0.207\pm0.071$ for short and long period ($\log [P] > 1.0$) Cepheids, respectively. This change of slope
is marginally significant using the $F$-test described in Paper I. One motivation for this paper is to investigate the physics behind PC
relations at maximum light and in particular why Galactic and LMC Cepheids display flat maximum light PC relations and SMC Cepheids do not.

Since the PC and PL relations are really projections of the period-luminosity-colour (PLC) relation, PC and PL relations are closely connected: changes in one are generally reflected in changes in the other. Hence one way to look for changes in Cepheid PL relations is to look for changes in PC and AC relations. Another important driving force behind this series of papers is to investigate the possible physics behind recent results
suggesting that the LMC (Large Magellanic Cloud) PL (and PC) relation is non-linear: in the sense that current LMC Cepheid data
are statistically consistent with two PL (and PC) relations of significantly differing slope with a break at/around a period of 10 days.
The observational evidence for this is presented in \citet{tam02}, Paper I \& III, \citet{nge05}, \citet{san04}
and \citet[][hereafter Paper IV]{nge06a}. It is also interesting to note, as has been pointed out in previous work, that the ``break'' period of $\log (P) \sim 1.0$ is also the location of the well known Cepheid pulsation resonance.
Paper IV use the OGLE \citep[Optical Gravitational Lensing Experiment,][]{uda99b} LMC data and phase all Cepheids to a common epoch and then plot
 both PC and PL relations as a function
of phase. This approach presents clear and compelling evidence of a break at a phase around 0.8 and also attest to the dynamic nature of the
PL relation. The PC relation follows changes in the PL relation very closely. However the standard PL relation used in the literature is at mean light: the average over phase of the multiphase relations presented in Paper IV. Because the
PL relation is indeed linear at certain phases (e.g., at maximum light), the net result is that the strength of the non-linearity is diluted
at mean light. We emphasize, though, that a battery of parametric and non-parametric statistical tests carried out on the LMC PL
relation at mean light in optical bands confirm the existence of the non-linearity at high confidence levels - greater than $99\%$.
These tests include the $F$-test, robust methods like Tukey's bi-weight function and non-parametric techniques \citep[Paper I;][]{nge05}.
In contrast the Galactic and Small Magellanic Cloud (SMC) PL and PC relations are found to be linear with current data \citep{uda99a,tam03,kan04,nge04}.
Since the Galaxy, LMC and SMC have different metallicity, it has been conjectured that metallicity is the key to
understanding this behavior.

We now briefly discuss some criticisms of results implying a non-linear LMC PL relation. One of these is that the OGLE LMC sample lacks sufficient long period Cepheids. Despite the fact that the $F$-test takes into account 
both the number and nature of the observations, \citet{nge06b} augmented the OGLE sample with data taken from, for example, \citet{cal91} and \citet{seb02}.
They conclude on the basis of their rigorous statistical test (the $F$-test) that even with this augmented
LMC Cepheid data, the LMC PL relation is still found to be non-linear in the sense described above (even though the difference in the slopes of the long and short period PL relations may be small).
Another criticism is that extinction errors are influencing the result even though only published extinction values are used.
Well established observations and theory have shown that the Galactic and LMC Cepheid PC relations are flat at maximum light, at least
for the longer period ($\log [P] > 1$) Cepheids \citep[see Paper I, II, III and][]{sim93}. 
If the extinction values required to make the mean light LMC PC and PL relations linear are adopted, then the LMC PC relation at maximum light is no longer flat. In this scenario, LMC Cepheids get hotter at maximum light as the period increases, which is in distinct disagreement with the theory and observations at maximum light \citep{sim93}. The referee has pointed out that if the amplitude increases sufficiently fast with period, then it could produce hotter temperatures at maximum light with increasing period. But note that the amplitude increases fairly rapidly after $10$ days and this is precisely when the observed PC relation is flattest. Further if the LMC Cepheids are such that they got hotter at maximum light as the period increases this is different to Galactic Cepheids at maximum light. We therefore believe that extinction errors are not the cause of the observed non-linear LMC PL relation. More detailed discussions regarding extinction errors are given in \citet{san04}, \citet{nge05} and \citet{nge06b}, and will not be repeated here. 

Since the Cepheid PL relation is a cornerstone of modern astrophysics, a proper understanding of its properties and the physics behind it is crucial. For example, it is of fundamental importance in
establishing an extra-galactic distance scale that is independent of CMB studies. An appropriate refinement of the
Cepheid PL distance scale, can provide a value of Hubble's constant accurate to less than 3-4\%. Though the current
Cepheid based estimate of $H_0$ is accurate to $10\%$ \citep{fre01}, a more accurate ($<5\%$) direct
measurement of $H_0$ via a Cepheid distance scale is still very relevant because it will help to break the
degeneracy between ${\Omega}_\mathrm{matter}$ and $H_0$ present from $WMAP$ CMB estimates \citep[see, for example,][]{teg04,fre05,hu05,spe07}.
CMB measurements can only estimate ${\Omega}_\mathrm{matter}H_0^2$ and estimate $H_0$ only if a simple flat
${\Lambda}$CDM model is assumed. Table 2 of \citet{spe07} points to the fact that an independent estimate of $H_0$ accurate to $1\%$ will
result in a reduction of the $65\%$ confidence interval on ${\Omega}_\mathrm{matter}$ by almost a factor of
two over that with $WMAP$ alone. This is a greater reduction than if other data such as 2dFGRS are used \citep[table 5 of][]{spe07}.
With the possibility of $Gaia$ and $SIM$ satellites producing more accurate parallaxes for many Cepheids a considerably more
accurate Cepheid zero point is well within reach. Better refinement of the slope of PL relation will become vital.
Equally as important, modeling the Cepheid PL relation can yield important information about the mass-luminosity (ML) relation
obeyed by Cepheids. Since these ML relations are sensitive functions of input physics, such as the amount of convective overshoot, 
such modeling is also important for theories of stellar evolution and pulsation.

A working theoretical hypothesis to explain the non-linear LMC PL and PC relation is the interaction of the HIF with the
stellar photosphere at certain
phases.
Paper II \& III published theoretical pulsation models of Galactic and LMC Cepheids, respectively, to model
this behavior. In these papers we found that the interaction of the HIF and stellar photosphere may play an important role in explaining the observed PC and PL behavior. We briefly summarize this theory in what follows (see Paper II \& III for more details). In a pulsating Cepheid, the HIF and stellar photosphere independently move in and out in the mass distribution as the star pulsates.   
In certain circumstances, the stellar photosphere can be located at the base of the HIF. In this case, when the the density is low the temperature
of the stellar photosphere and hence the effective temperature of the Cepheid at that phase experiences a much reduced
dependence on pulsation period. Thus the PC relation is flatter in this situation. Because the PL relation is intimately connected to
changes in the PC relation, the PL relation slope will also change.
Because this HIF-photosphere engagement happens suddenly, the changes in the PC and PL relation are sharp.
The phase and period at which such a low density HIF-photosphere interaction can occur depends on the ML relation since this controls the location of the HIF in the mass distribution \citep{kan95,kan96}. The connection to
metallicity occurs because the ML relation, taken from stellar evolution calculations, is indeed a function of metallicity.

In this paper we extend the work of Paper II \& III by constructing a large grid of full amplitude hydrodynamic models of Cepheids with an 
SMC metallicity and study how
the interaction of the photosphere and HIF affects the PC (and hence the PL relation) in these models compared to previous work looking at Galactic (Paper II) and
LMC (Paper III) models.
Section 2 outlines the methodology and pulsation codes we used in this paper and Section 3 presents our results for the SMC models. The conclusion and discussion is given in Section 4.

     \begin{table}
       \centering
       \caption{Input parameters for SMC Cepheid models with periods obtained from a linear analysis. The periods, $P_0$ and $P_1$, are referred to the fundamental and first overtone periods, respectively. Similarly for the growth rate, $\eta$. Both of the mass and luminosity are in Solar units, the temperature is in K and the period is in days.}
       \label{tabinput}
       \begin{tabular}{ccccccc} \hline
         $M$ & $\log(L)$ & $T_{eff} $ & $P_0$ & $\eta_0$ & $P_1$ & $\eta_1$ \\
         \hline 
         \multicolumn{7}{c}{ML Relation from \citet{bon00}} \\
         12.0 & 4.612 & 5440 & 52.695 & 0.247 & 32.94 & $-0.088$ \\
	 10.8 & 4.458 & 5350 & 44.130 & 0.249 & 27.97 & $-0.029$ \\
         9.50 & 4.272 & 5230 & 35.981 & 0.175 & 23.01 & $-0.041$ \\
	 8.70 & 4.144 & 5180 & 30.519 & 0.110 & 19.68 & $-0.057$ \\
	 8.00 & 4.022 & 5320 & 22.562 & 0.106 & 15.09 & $-0.016$ \\
	 7.10 & 3.848 & 5340 & 16.965 & 0.065 & 11.53 & $-0.020$ \\
	 6.50 & 3.720 & 5410 & 13.219 & 0.050 & 9.129 & $-0.011$ \\
	 5.90 & 3.579 & 5450 & 10.363 & 0.033 & 7.226 & $-0.013$ \\
	 5.80 & 3.554 & 5470 & 9.8372 & 0.032 & 6.879 & $-0.010$ \\
	 5.50 & 3.476 & 5500 & 8.5701 & 0.026 & 6.022 & $-0.010$ \\
         5.20 & 3.395 & 5530 & 7.4236 & 0.020 & 5.238 & $-0.010$ \\
	 5.10 & 3.367 & 5530 & 7.1171 & 0.018 & 5.023 & $-0.012$ \\
	 5.00 & 3.338 & 5550 & 6.7222 & 0.017 & 4.755 & $-0.011$ \\
	 4.80 & 3.278 & 5570 & 6.0683 & 0.014 & 4.301 & $-0.012$ \\
	 4.60 & 3.217 & 5600 & 5.4218 & 0.012 & 3.854 & $-0.011$ \\
         \multicolumn{7}{c}{ML Relation from \citet{chi89}} \\
         8.50 & 4.504 & 5290 & 62.455 & 0.504 & 35.31 & $-0.230$ \\
	 7.80 & 4.384 & 5310 & 50.525 & 0.441 & 29.57 & $-0.151$ \\
	 7.00 & 4.232 & 5330 & 39.008 & 0.368 & 23.67 & $-0.088$ \\
	 6.50 & 4.129 & 5370 & 32.048 & 0.316 & 19.96 & $-0.057$ \\
	 5.70 & 3.945 & 5410 & 23.308 & 0.242 & 15.02 & $-0.023$ \\
	 4.70 & 3.675 & 5430 & 15.163 & 0.144 & 10.09 & $-0.017$ \\
	 4.30 & 3.551 & 5470 & 12.177 & 0.114 & 8.234 & $-0.012$ \\
	 3.95 & 3.432 & 5490 & 10.029 & 0.082 & 6.852 & $-0.016$ \\
	 3.80 & 3.378 & 5520 & 9.0437 & 0.074 & 6.221 & $-0.012$ \\
	 3.60 & 3.302 & 5550 & 7.9025 & 0.062 & 5.475 & $-0.011$ \\
	 3.45 & 3.243 & 5550 & 7.2318 & 0.050 & 5.022 & $-0.016$ \\
	 3.40 & 3.222 & 5570 & 6.9154 & 0.049 & 4.817 & $-0.013$ \\
	 3.30 & 3.181 & 5590 & 6.4078 & 0.045 & 4.479 & $-0.013$ \\
	 3.20 & 3.138 & 5610 & 5.9276 & 0.040 & 4.158 & $-0.012$ \\
	 3.00 & 3.047 & 5640 & 5.0804 & 0.031 & 3.581 & $-0.014$ \\
         \hline 
       \end{tabular}
     \end{table}

     \begin{table}
       \centering
       \caption{Temperatures at maximum and minimum light from full-amplitude non-linear model calculations. The periods, luminosity and temperature are in days, $L_{\odot}$ and K, respectively. \label{tabmaxmin}}
       \begin{tabular}{ccccc} \hline
         $P$ & $L_{max}$ & $T_{max}$ & $L_{min}$ & $T_{min}$ \\ 
         \hline 
         \multicolumn{5}{c}{ML Relation from \citet{bon00}} \\ 
         52.695 & 53454.42 & 5988.56 & 27021.41 & 5063.31 \\
	 44.130 & 35895.77 & 5826.30 & 19377.45 & 5022.77 \\
         35.981 & 21806.85 & 5480.18 & 14111.82 & 5031.24 \\
	 30.519 & 15662.90 & 5380.53 & 11737.37 & 5095.86 \\
	 22.562 & 11969.05 & 5626.25 & 8755.998 & 5063.77 \\
	 16.965 & 7880.140 & 5581.23 & 6253.112 & 5159.75 \\
	 13.219 & 5731.252 & 5560.19 & 4789.290 & 5260.59 \\
	 10.363 & 3934.212 & 5488.40 & 3546.536 & 5359.55 \\
	 9.837  & 3699.569 & 5495.11 & 3336.452 & 5378.74 \\
	 8.5701 & 3093.723 & 5609.36 & 2808.320 & 5421.22 \\
	 7.4236 & 2555.649 & 5607.43 & 2357.080 & 5471.61 \\
	 7.1171 & 2389.147 & 5605.87 & 2225.435 & 5481.90 \\
	 6.7222 & 2234.535 & 5625.98 & 2089.173 & 5507.61 \\
	 6.0683 & 1943.932 & 5642.11 & 1839.759 & 5538.16 \\
	 5.4218 & 1681.461 & 5664.88 & 1604.783 & 5574.01 \\
         \multicolumn{5}{c}{ML Relation from \citet{chi89}} \\ 
        62.455 & 37223.13 & 5497.13 & 20783.42 & 4867.51 \\
	50.525 & 28606.35 & 5590.89 & 15062.32 & 4869.70 \\
	39.008 & 20384.64 & 5630.58 & 10622.78 & 4916.65 \\
	32.048 & 16268.92 & 5710.44 & 8633.998 & 5001.55 \\
	23.308 & 10611.59 & 5727.19 & 6356.127 & 5179.16 \\
	15.163 & 5419.509 & 5562.85 & 3864.655 & 5147.97 \\
	12.177 & 4024.093 & 5771.01 & 2991.331 & 5226.61 \\
	10.029 & 3020.997 & 5703.97 & 2380.306 & 5291.96 \\
	9.0437 & 2653.422 & 5727.42 & 2129.332 & 5341.58 \\
        7.9025 & 2192.885 & 5700.61 & 1826.794 & 5404.38 \\
        7.2318 & 1872.266 & 5655.25 & 1622.456 & 5433.35 \\
        6.9154 & 1774.033 & 5653.59 & 1550.829 & 5465.08 \\
        6.4078 & 1592.429 & 5645.12 & 1409.870 & 5495.35 \\
        5.9276 & 1430.522 & 5645.05 & 1274.321 & 5516.67 \\
        5.0804 & 1148.483 & 5757.16 & 1041.914 & 5559.16 \\
	 \hline 
       \end{tabular}
     \end{table}

\section{Methods and SMC Models}

The codes for computational and numerical methods for doing the pulsation modelings are described in \citet{yec98} and \citet{kol02}, and are exactly the same as used
in Paper II \& III. In brief, the input parameters to the pulsation codes include the mass ($M$), luminosity ($L$), effective temperature ($T_{eff}$) and chemical composition ($X,Z$). In this paper, the chemical composition is set to be $(X,Z)=(0.70,0.004)$ to represent the SMC hydrogen and metallicity abundance (by mass). The input effective temperatures are chosen to ensure the models oscillate in the fundamental mode and located within the Cepheid instability strip. To be consistent with Paper II \& III, we adopt the same ML relations as given below:

     \begin{enumerate}   
     \item ML relation given in \citet{bon00}:
       \begin{eqnarray}
         \log(L) & = & 0.90 + 3.35\log(M) + 1.36\log(Y) - 0.34\log(Z), \nonumber \\
         & = & 3.35\log(M) + 0.996.
       \end{eqnarray}
     
     \item ML relation given in \citet{chi89} for $Z=0.001$:
       \begin{eqnarray}
         \log(L) & = & 3.22\log(M)+1.511.
       \end{eqnarray}
     \end{enumerate}

\ni  The units for both $M$ and $L$ are in Solar units. To convert the temperatures from the models to the $(V-I)$ colours, we use the {\tt BaSeL} atmosphere database \citep{lej02,wes02} to construct a fit giving temperature and effective gravity (see Paper II) as a function of $(V-I)$ colour. The bolometric corrections ($BC$) are converted in a similar manner. Note that the effect of a micro-turbulence parameter that varies with phase has been discussed in Paper III: its variation with phase does not influence the results presented in this paper. To convert the observed colours to the temperatures appropriate for the SMC data, we use the prescriptions given in \citet{bea01}: 

       \begin{eqnarray}
         \log(g) & = & 2.62 - 1.21 \log (P), \nonumber \\
         \log(T_{eff}) & = & 3.91611 + 0.0055\log(g) - 0.2482(V-I)_0, \nonumber \\
         \Delta T & = & \log(T_{eff}) - 3.772, \nonumber \\
         BC & = & -0.0324 + 2.01\Delta T - 0.0217\log(g)  \nonumber \\
            &   & - 10.31(\Delta T)^2.\nonumber
       \end{eqnarray}

\ni More details of the methodology, the pulsation codes and the temperature-colour conversions can be found in Paper II \& III, and will not be repeated in detail here.

\section{Results}

     \begin{table*}
       \centering
       \caption{Temperatures at mean light from full-amplitude non-linear model calculations. See Paper II for the meanings of $<L>$, $L_{mean}$, $T_{mean}$ and $T^{inter}_{mean}$. The periods, luminosity and temperature are in days, $L_{\odot}$ and K, respectively.}
       \label{tabmean}
       {\footnotesize
       \begin{tabular}{cccccccc} \hline
         $P$ & $<L>$ & $L_{mean}(asc)$ & $T_{mean} (asc)$ & $L_{mean} (des)$ & $T_{mean} (des)$ & $T_{mean}^{inter}$ (asc) & $T_{mean}^{inter}$ (des)\\ 
         \hline 
         \multicolumn{8}{c}{ML Relation from \citet{bon00}} \\
         52.695 & 40955.18 & 41381.50 & 5794.90 & 40910.47 & 5191.18 & 5779.38 & 5192.37 \\
         44.130 & 29325.71 & 29139.10 & 5704.73 & 29292.77 & 5107.24 & 5713.09 & 5108.67 \\
         35.981 & 19241.07 & 19236.72 & 5557.14 & 19209.26 & 5025.52 & 5557.48 & 5027.67 \\
	 30.519 & 13945.37 & 13879.68 & 5404.28 & 13960.74 & 4992.63 & 5411.22 & 4991.33 \\
	 22.562 & 10490.84 & 10518.15 & 5535.73 & 10507.93 & 5138.62 & 5532.01 & 5136.74 \\
	 16.965 & 7033.900 & 7014.821 & 5486.05 & 7035.167 & 5207.44 & 5489.46 & 5207.20 \\
	 13.219 & 5237.352 & 5240.488 & 5502.74 & 5232.986 & 5324.74 & 5502.19 & 5326.06 \\
	 10.363 & 3787.302 & 3787.677 & 5513.59 & 3791.380 & 5417.43 & 5513.40 & 5415.87 \\
	 9.8372 & 3576.717 & 3579.774 & 5541.08 & 3572.195 & 5431.94 & 5539.47 & 5433.71 \\
	 8.5701 & 2993.448 & 2996.177 & 5570.00 & 2991.997 & 5465.07 & 5568.28 & 5465.71 \\
	 7.4236 & 2480.871 & 2483.940 & 5593.50 & 2482.250 & 5501.82 & 5591.18 & 5501.01 \\
	 7.1171 & 2324.563 & 2325.461 & 5586.57 & 2324.964 & 5506.19 & 5585.84 & 5505.89 \\
	 6.7222 & 2175.275 & 2175.738 & 5605.57 & 2175.519 & 5526.88 & 5605.17 & 5526.69 \\
	 6.0683 & 1897.128 & 1899.297 & 5620.35 & 1897.321 & 5553.28 & 5618.19 & 5553.09 \\
	 5.4218 & 1644.994 & 1643.720 & 5641.55 & 1644.522 & 5587.32 & 5642.97 & 5587.89 \\
         \multicolumn{8}{c}{ML Relation from \citet{chi89}} \\  
         62.455 & 31923.84 & 31934.40 & 5622.99 & 31788.30 & 5080.26 & 5622.53 & 5084.43 \\
	 50.525 & 24193.87 & 24220.00 & 5657.21 & 24125.18 & 5092.35 & 5655.55 & 5094.78 \\
	 39.008 & 17315.32 & 17452.68 & 5718.16 & 17393.96 & 5111.54 & 5706.24 & 5107.63 \\
	 32.048 & 13480.61 & 13231.90 & 5702.29 & 13467.68 & 5118.54 & 5729.31 & 5119.08 \\
	 23.308 & 9065.182 & 9050.282 & 5778.65 & 9095.283 & 5184.01 & 5780.89 & 5179.84 \\
	 15.163 & 4886.975 & 4876.629 & 5719.32 & 4886.911 & 5268.85 & 5722.14 & 5268.87 \\
	 12.177 & 3545.551 & 3517.459 & 5664.05 & 3535.748 & 5288.78 & 5674.57 & 5291.86 \\
	 10.029 & 2700.579 & 2688.289 & 5642.62 & 2708.607 & 5350.09 & 5648.19 & 5346.47 \\
	 9.0437 & 2383.037 & 2371.188 & 5644.71 & 2377.980 & 5391.28 & 5650.44 & 5394.08 \\
	 7.9025 & 2003.229 & 1999.609 & 5639.31 & 2002.772 & 5459.24 & 5641.05 & 5459.58 \\
	 7.2318 & 1746.108 & 1747.221 & 5620.58 & 1745.418 & 5640.07 & 5620.23 & 5640.65 \\
	 6.9154 & 1666.125 & 1666.384 & 5630.74 & 1666.937 & 5651.07 & 5630.80 & 5649.58 \\
	 6.4078 & 1514.393 & 1513.263 & 5663.32 & 1516.365 & 5549.10 & 5664.74 & 5547.18 \\
	 5.9276 & 1371.416 & 1375.426 & 5695.05 & 1373.677 & 5570.67 & 5689.58 & 5568.37 \\
	 5.0804 & 1113.786 & 1113.660 & 5717.14 & 1115.086 & 5603.50 & 5717.35 & 5602.00 \\
         \hline 
       \end{tabular}
       }
     \end{table*} 


     \begin{figure*}
       \vspace{0cm}
       \hbox{\hspace{1.2cm}\epsfxsize=7.5cm \epsfbox{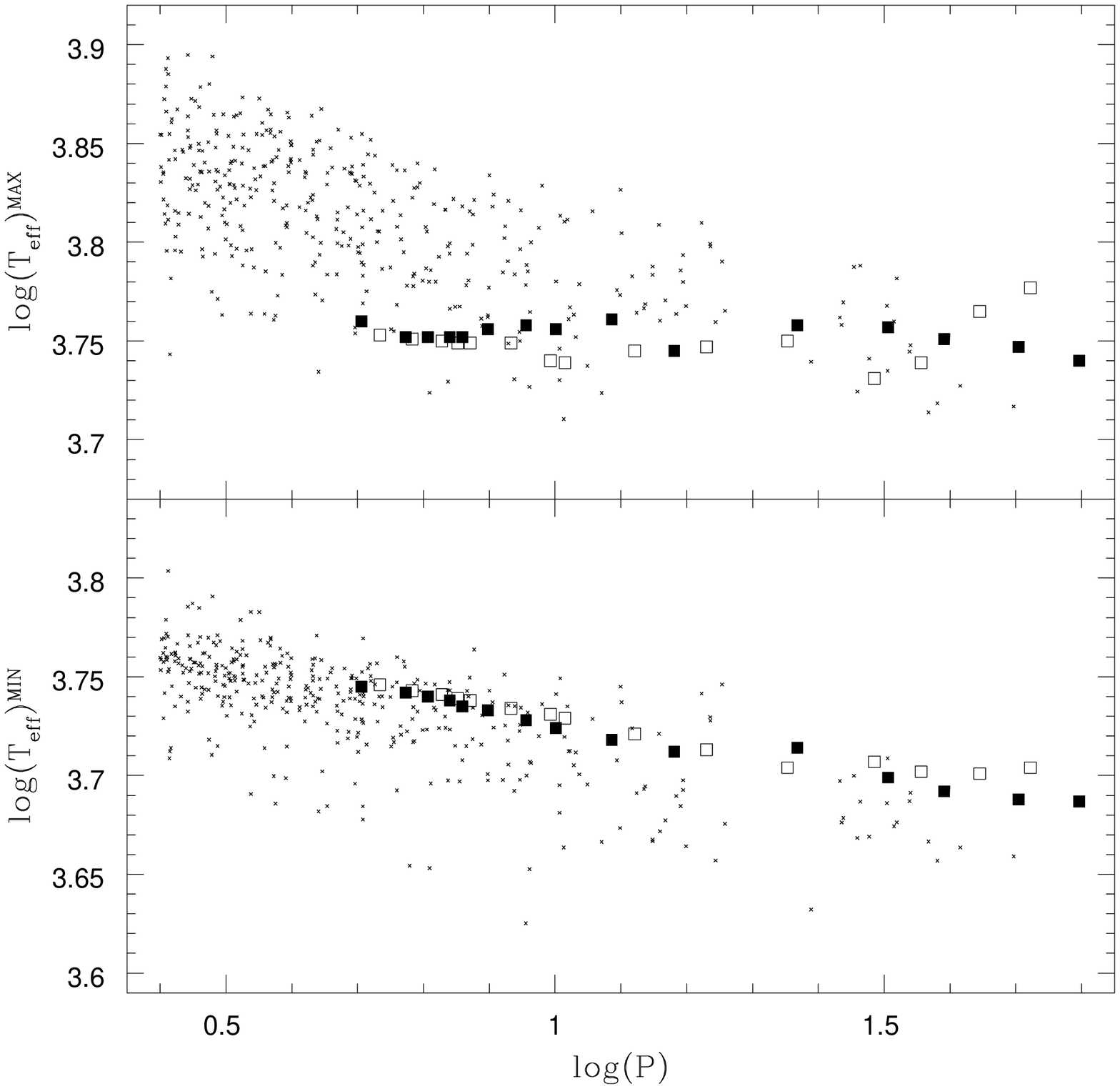}
         \epsfxsize=7.5cm \epsfbox{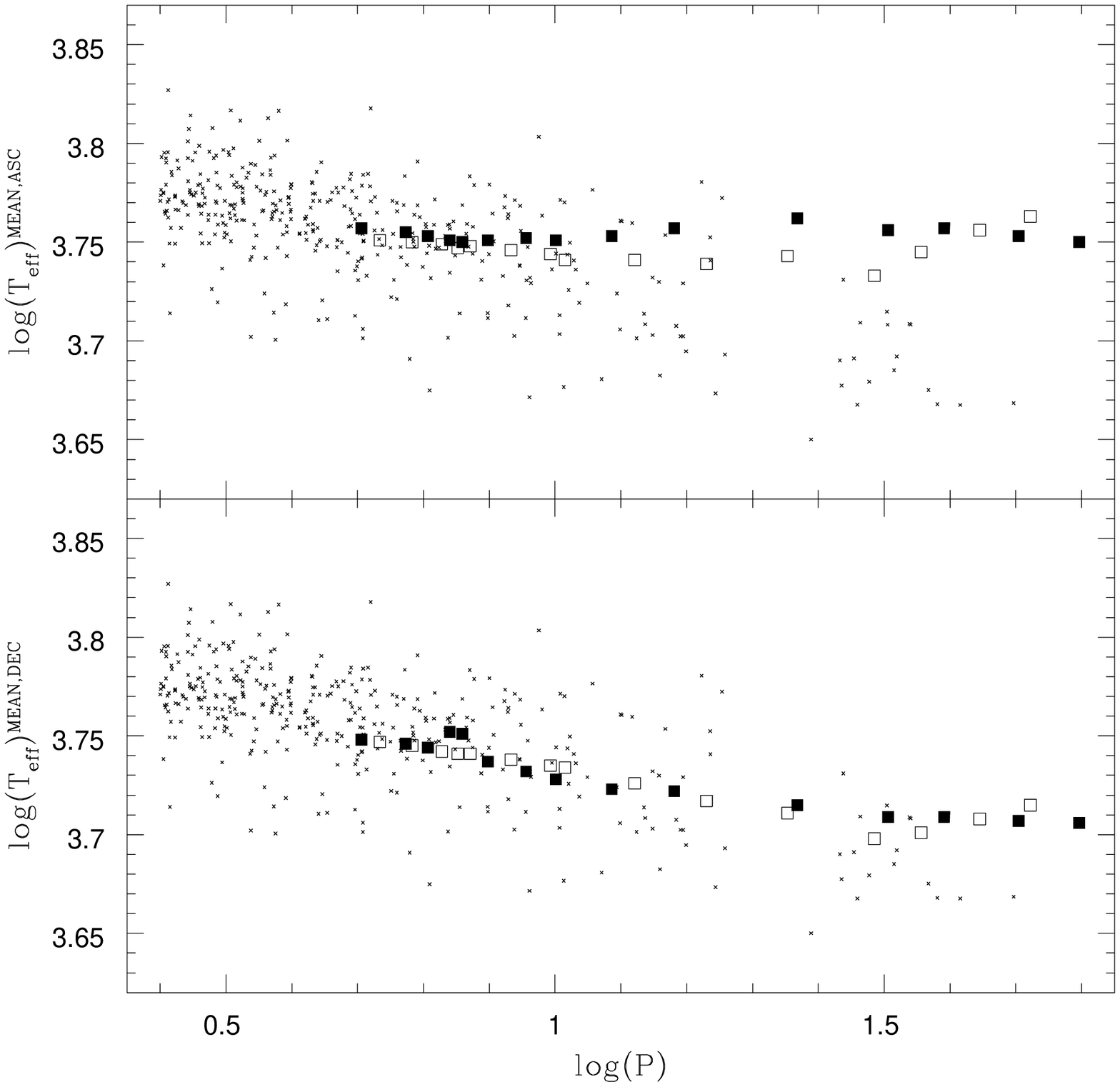}}
       \vspace{0cm}
       \caption{The plots of $\log(T)$-$\log(P)$ relations for the SMC data (small crosses) and models. The open and solid squares are for the models calculated with \citet{bon00} and \citet{chi89} ML relations, respectively. The conversion of the $(V-I)$ colours to the temperature are done using the equations given in \citet{bea01}. Left panel: $\log(T)$-$\log(P)$ relations at maximum and minimum light. Right panel: $\log(T)$-$\log(P)$ relations at mean light for both of the ascending and descending means. }
       \label{pt}
     \end{figure*}


     \begin{figure*}
       \vspace{0cm}
       \hbox{\hspace{1.2cm}\epsfxsize=7.5cm \epsfbox{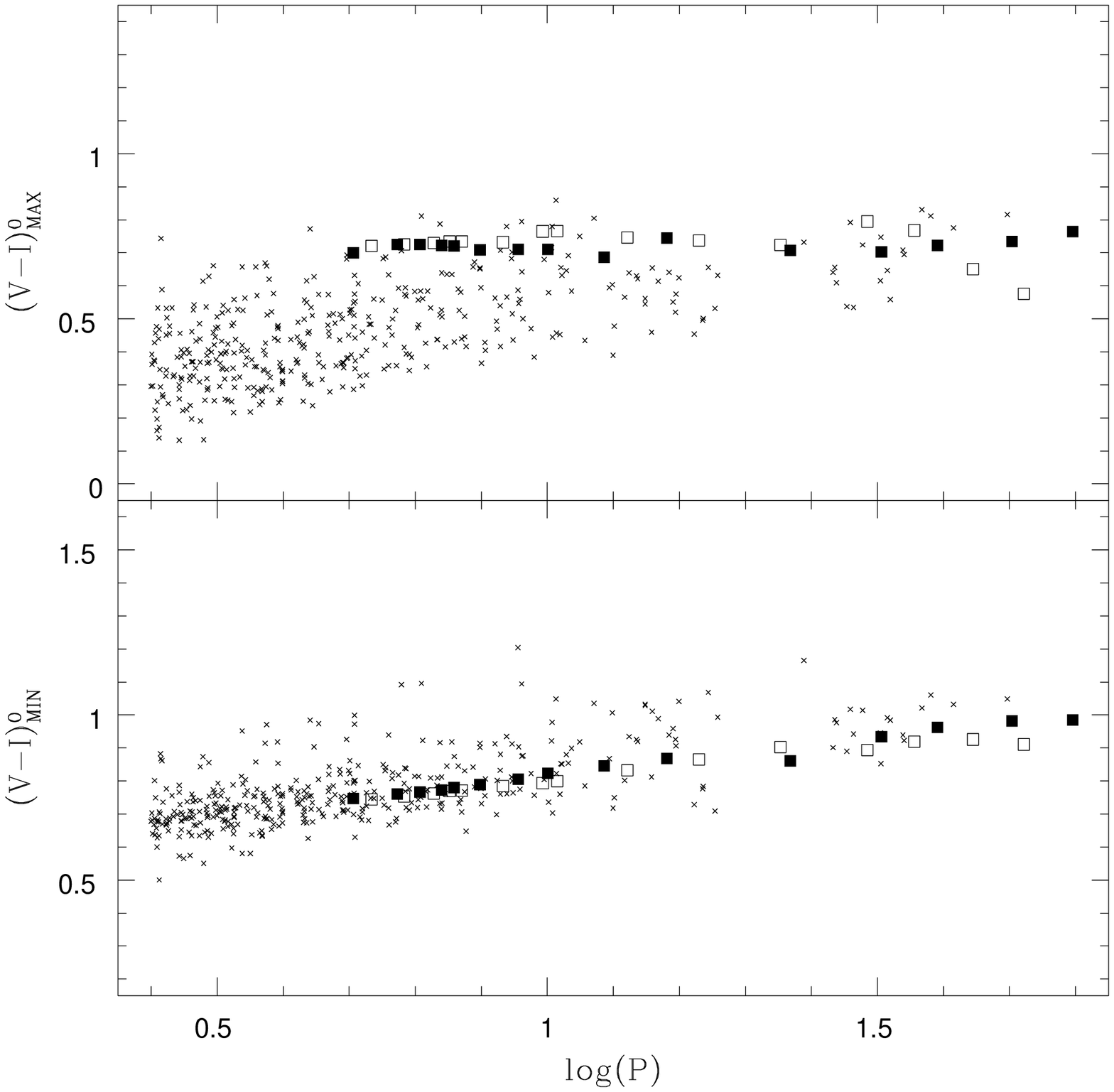}
         \epsfxsize=7.5cm \epsfbox{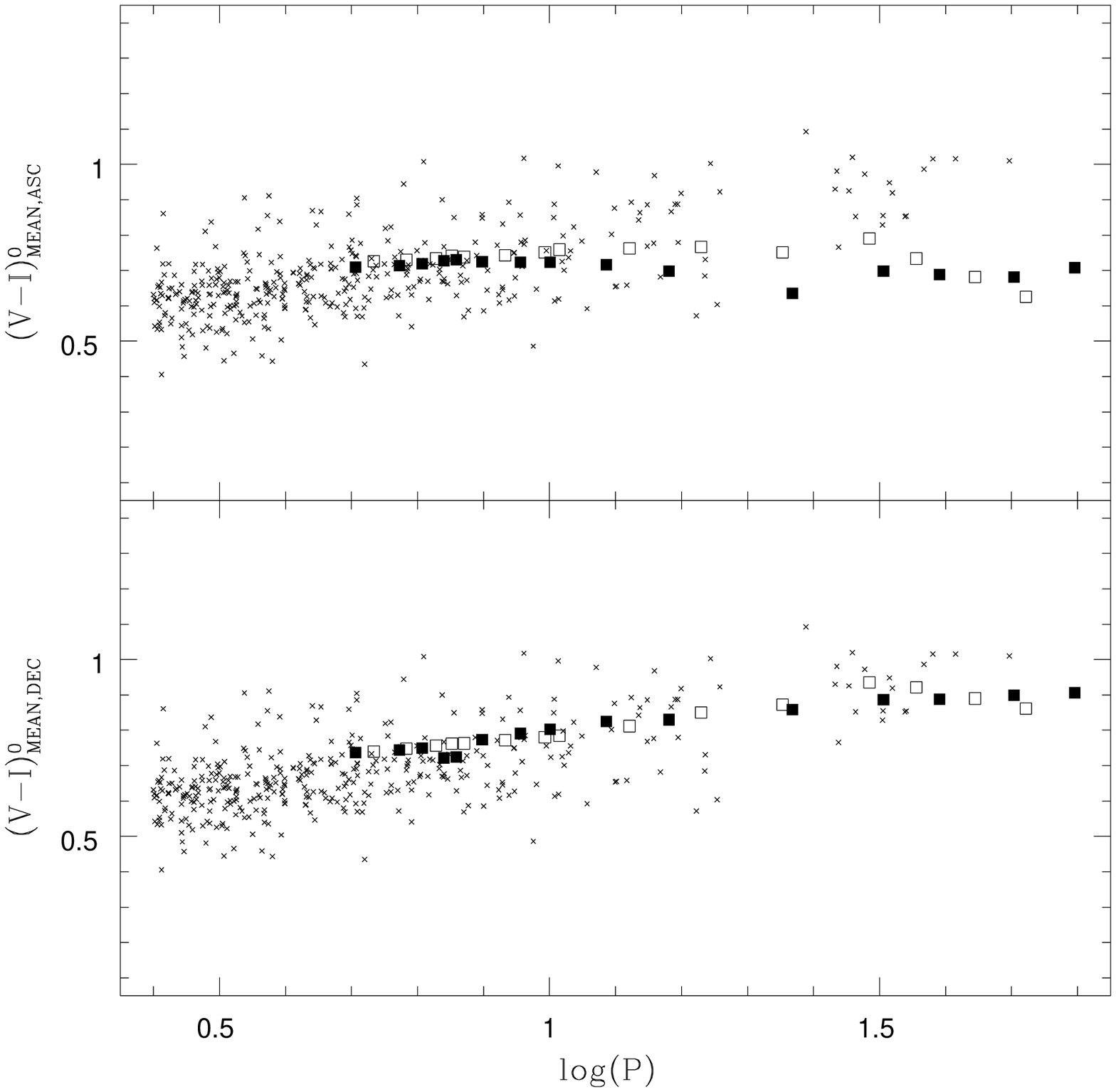}}
       \vspace{0cm}
       \caption{The SMC PC relations with the results from the models. The symbols are the same as in Figure \ref{pt}. The temperatures of the models are converted to the $(V-I)$ colour using the {\tt BaSeL} database. Left panel: PC relations at maximum and minimum light. Right panel: PC relations at mean light for both of the ascending and descending means.}
       \label{pc}
     \end{figure*}


     \begin{figure*}
       \vspace{0cm}
       \hbox{\hspace{1.2cm}\epsfxsize=7.5cm \epsfbox{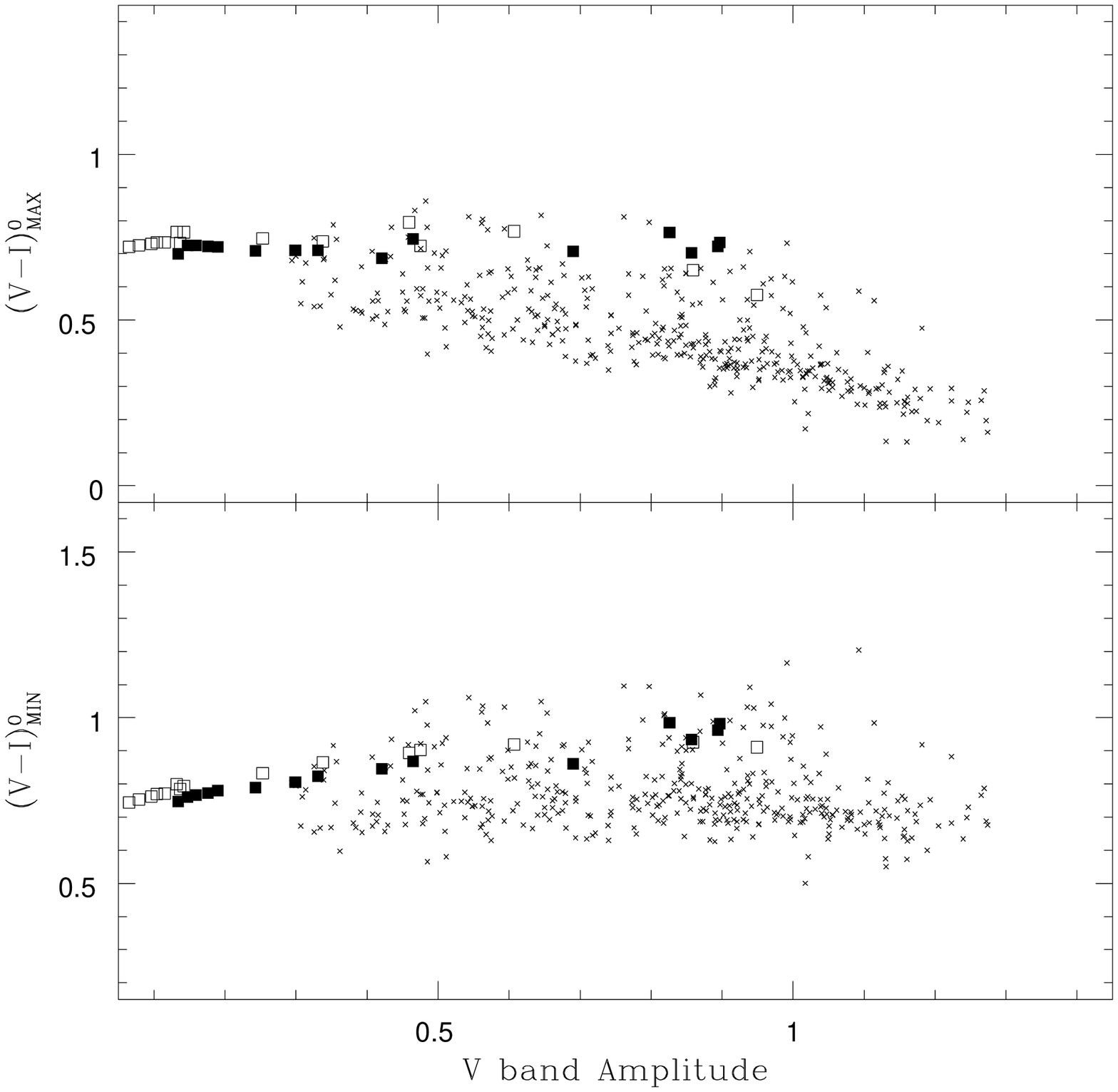}
         \epsfxsize=7.5cm \epsfbox{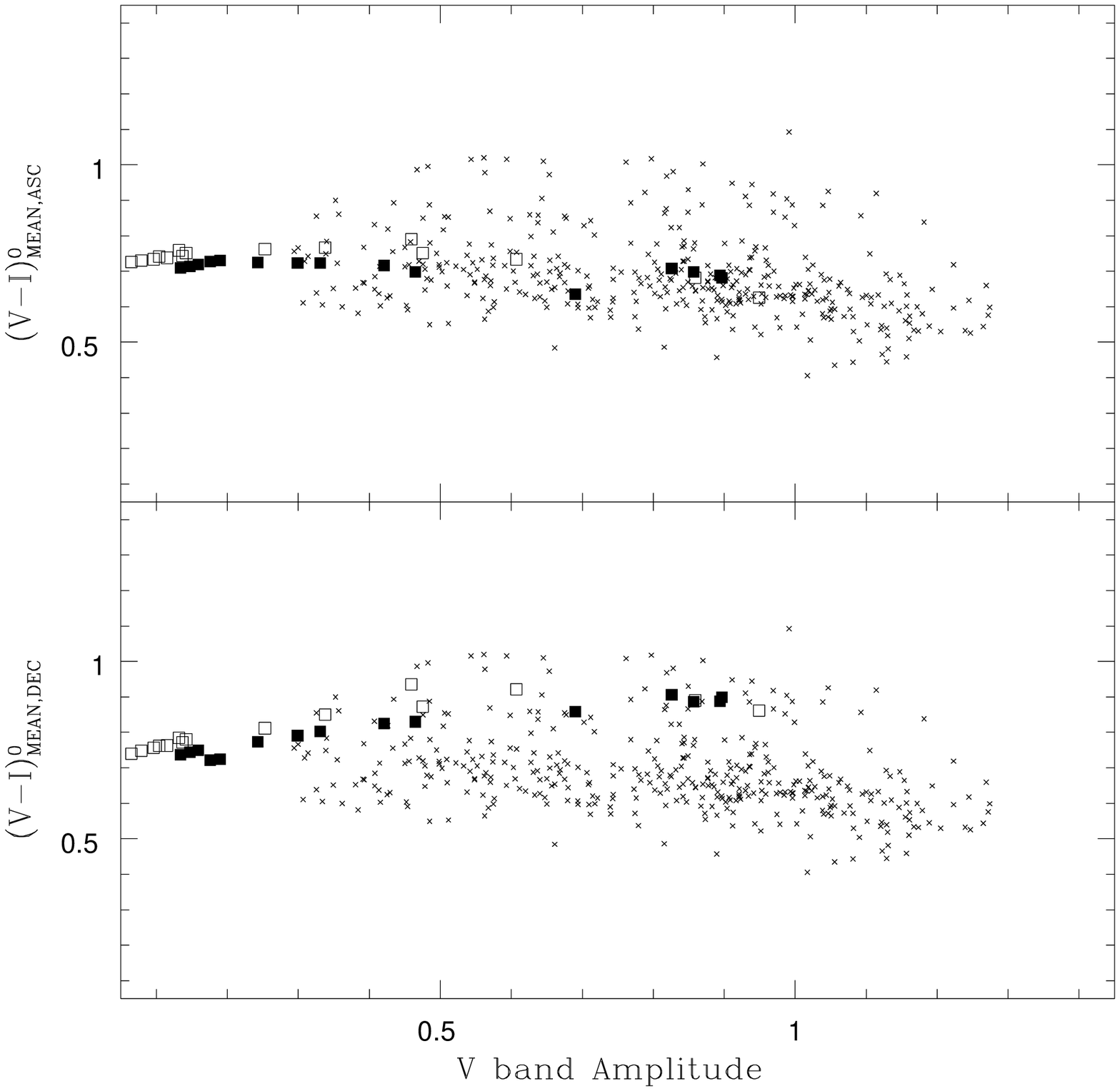}}
       \vspace{0cm}
       \caption{The SMC AC relations with the results from the models. The symbols are the same as in Figure \ref{pt}. The bolometric light curves from models are converted to $V$-band light curves with the $BC$ obtained from the {\tt BaSeL} database. Left panel: AC relations at maximum and minimum light. Right panel: AC relations at mean light for both of the ascending and descending means.}
       \label{ac}
     \end{figure*}

The results for our SMC models are collectively summarized in Table \ref{tabinput}-\ref{tabmean}.
Table \ref{tabinput} presents the input mass (and hence the luminosity) and effective temperature for our SMC models and the resulting periods from linear non-adiabatic calculations. Table \ref{tabinput} displays the fact that the models constructed in this study occupied a wide range of periods with an unstable and stable fundamental and first overtone mode respectively. Thus we avoid regions of the instability strip which are susceptible to first overtone or double mode pulsation. 
Table \ref{tabmaxmin} gives the temperatures at the maximum and minimum luminosity for the models calculated in Table \ref{tabinput}.
Table \ref{tabmean}, with identical layout as in table 3 of Paper II, presents more detailed information regarding photospheric temperatures at mean luminosities 
during the ascending and descending branches of the light (or luminosity) curves (see Paper II \& III for more details). 

       \begin{table*}
         \centering
         \caption{The SMC period-colour relation in the form of $(V-I)=a\log(P)+b$, and $\sigma$ is the dispersion of the relation. Long and short periods refer to Cepheids with $\log(P)>1.0$ and $\log(P)<1.0$, respectively. See Paper I \& III for the definition of the phases.}
         \label{tabpc}
         \begin{tabular}{lccccccccc} \hline
            & \multicolumn{3}{c}{All, $N=391$} & \multicolumn{3}{c}{Long period, $N=57$} & \multicolumn{3}{c}{Short period, $N=334$} \\  
	   Phase & $a_\mathrm{All}$ & $b_\mathrm{All}$ & $\sigma_\mathrm{All}$ & $a_\mathrm{Long}$ & $b_\mathrm{Long}$ & $\sigma_\mathrm{Long}$ & $a_\mathrm{Short}$ & $b_\mathrm{Short}$ & $\sigma_\mathrm{Short}$ \\
           \hline   
           Maximum & $0.324\pm0.021$ & $0.230\pm0.016$ & 0.114 & $0.207\pm0.071$ & $0.366\pm0.089$ & 0.106 & $0.396\pm0.039$ & $0.187\pm0.025$ & 0.114 \\
           Mean    & $0.264\pm0.015$ & $0.476\pm0.011$ & 0.079 & $0.280\pm0.060$ & $0.459\pm0.075$ & 0.090 & $0.241\pm0.026$ & $0.489\pm0.017$ & 0.077 \\
           Phmean  & $0.272\pm0.017$ & $0.491\pm0.013$ & 0.090 & $0.340\pm0.074$ & $0.414\pm0.092$ & 0.111 & $0.229\pm0.030$ & $0.517\pm0.019$ & 0.086 \\
           Minimum & $0.276\pm0.015$ & $0.575\pm0.012$ & 0.081 & $0.229\pm0.066$ & $0.635\pm0.082$ & 0.098 & $0.282\pm0.027$ & $0.571\pm0.017$ & 0.078 \\
           \hline
         \end{tabular}
       \end{table*}

       \begin{table*}
         \centering
         \caption{The SMC amplitude-colour relation in the form of $(V-I)=aV_{amp}+b$, and $\sigma$ is the dispersion of the relation. Long and short periods refer to Cepheids with $\log(P)>1.0$ and $\log(P)<1.0$, respectively. See Paper I \& III for the definition of the phases.}
         \label{tabac}
         \begin{tabular}{lccccccccc} \hline
            & \multicolumn{3}{c}{All, $N=391$} & \multicolumn{3}{c}{Long period, $N=57$} & \multicolumn{3}{c}{Short period, $N=334$} \\  
	   Phase & $a_\mathrm{All}$ & $b_\mathrm{All}$ & $\sigma_\mathrm{All}$ & $a_\mathrm{Long}$ & $b_\mathrm{Long}$ & $\sigma_\mathrm{Long}$ & $a_\mathrm{Short}$ & $b_\mathrm{Short}$ & $\sigma_\mathrm{Short}$ \\
           \hline 
           Maximum & $-0.438\pm0.021$ & $0.816\pm0.018$ & 0.099 & $-0.308\pm0.070$ & $0.848\pm0.053$ & 0.098 & $-0.419\pm0.018$ & $0.779\pm0.016$ & 0.081 \\
           Mean    & $-0.166\pm0.021$ & $0.799\pm0.018$ & 0.099 & $0.009\pm0.076$ & $0.800\pm0.057$ & 0.106 & $-0.152\pm0.017$ & $0.765\pm0.015$ & 0.078 \\
           Phmean  & $-0.158\pm0.023$ & $0.815\pm0.020$ & 0.110 & $0.034\pm0.092$ & $0.810\pm0.070$ & 0.130 & $-0.144\pm0.019$ & $0.779\pm0.017$ & 0.086 \\
           Minimum & $-0.065\pm0.023$ & $0.826\pm0.020$ & 0.109 & $0.104\pm0.076$ & $0.842\pm0.058$ & 0.107 & $-0.047\pm0.020$ & $0.787\pm0.017$ & 0.089 \\
           \hline
         \end{tabular}
       \end{table*}


\begin{figure*}
  \begin{tabular}{ccc}
    {\epsfxsize=5.5truecm \epsfysize=5.5truecm \epsfbox[17 144 590 715]{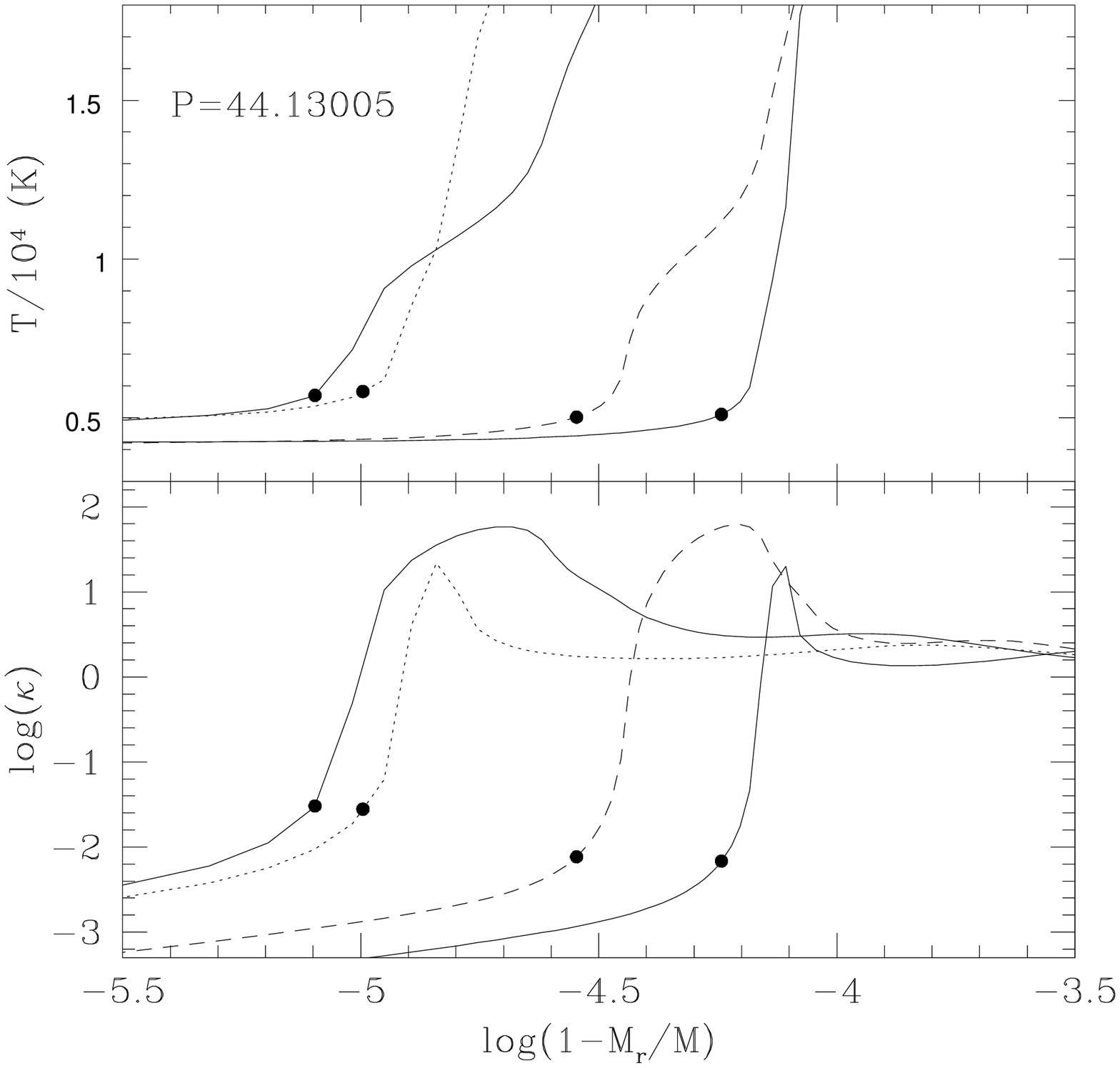}} &
    {\epsfxsize=5.5truecm \epsfysize=5.5truecm \epsfbox[17 144 590 715]{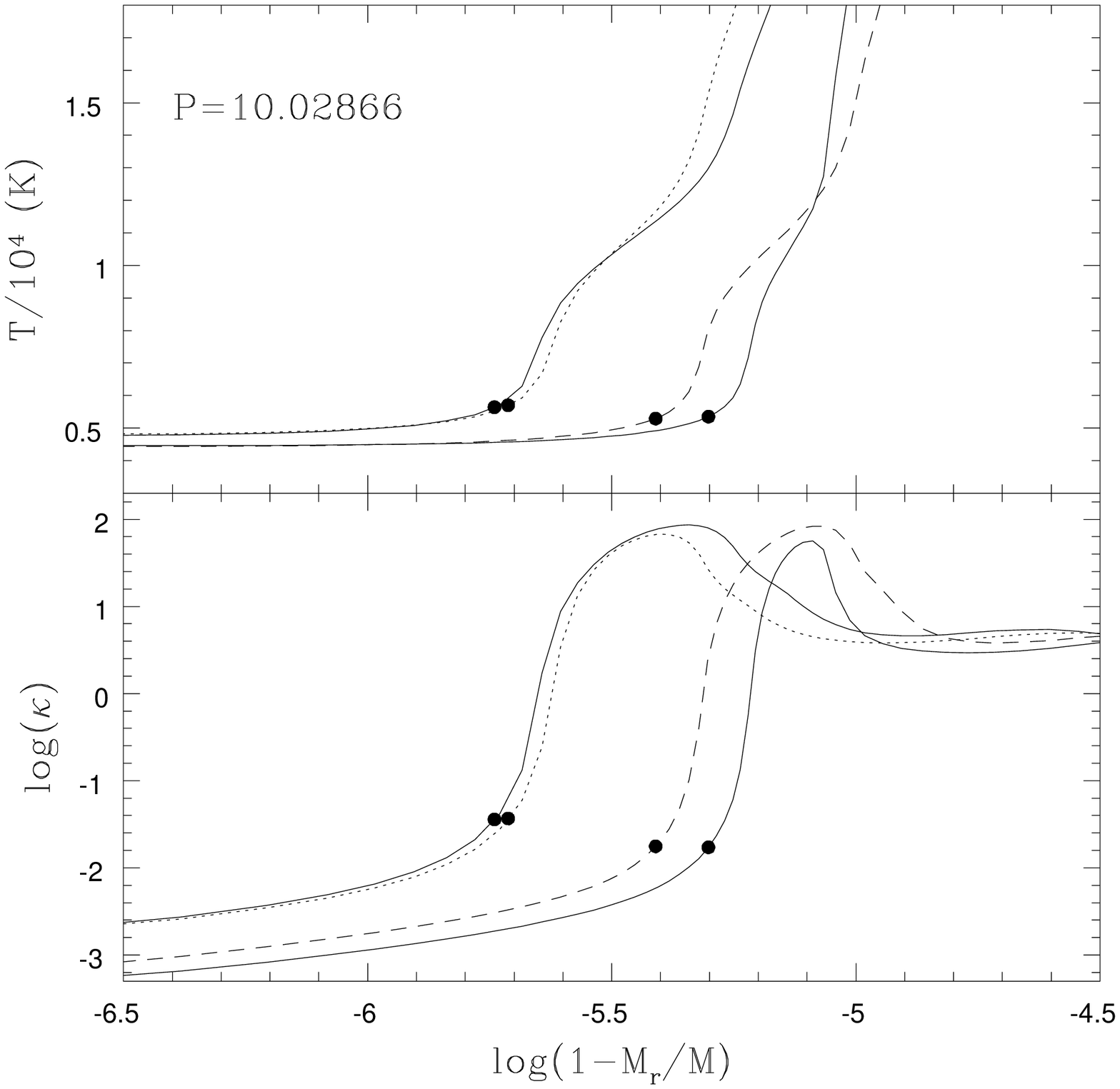}} &
    {\epsfxsize=5.5truecm \epsfysize=5.5truecm \epsfbox[17 144 590 715]{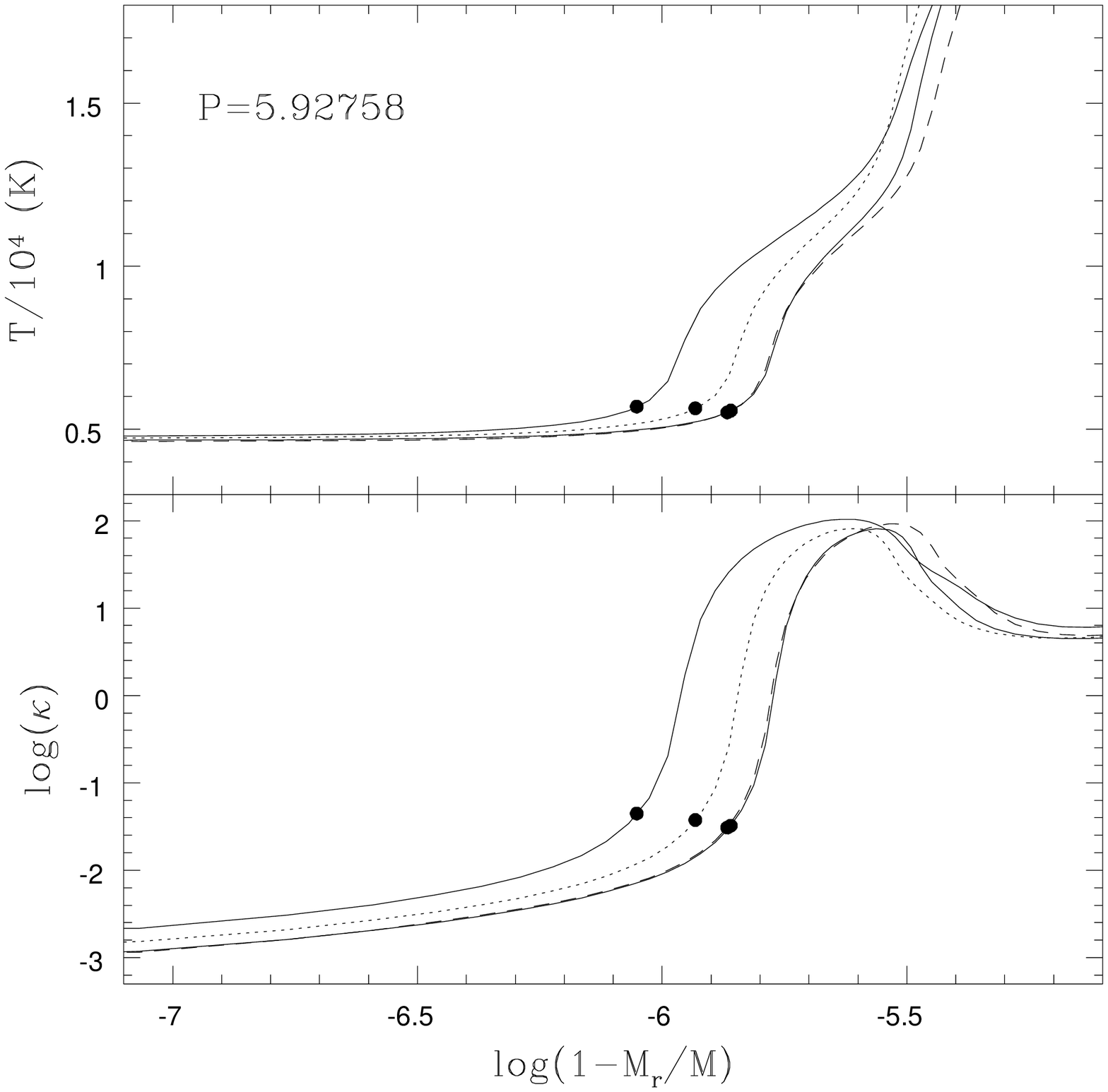}} \\
  \end{tabular}
  \caption{The temperature (top panels) and the opacity ($\kappa$, bottom panels) profiles, plotted in terms of the internal mass distribution ($\log[1-M_r/M]$, where $M_r$ is mass within radius $r$ and $M$ is the total mass), for a long period, a 10-days period and a short period SMC models. The dotted, solid and dashed curves are for the profiles at maximum, mean and minimum light, respectively. The filled circles denote the location of the photosphere at $\tau=2/3$ for each phases. The mean light profiles at the ascending and descending branch are the solid curves that lie close to the profiles at maximum light (dotted curves) and minimum light (dashed curves), respectively.}
  \label{profile}
\end{figure*}


     \begin{figure*}
       \vspace{0cm}
       \hbox{\hspace{1.2cm}\epsfxsize=7.5cm \epsfbox{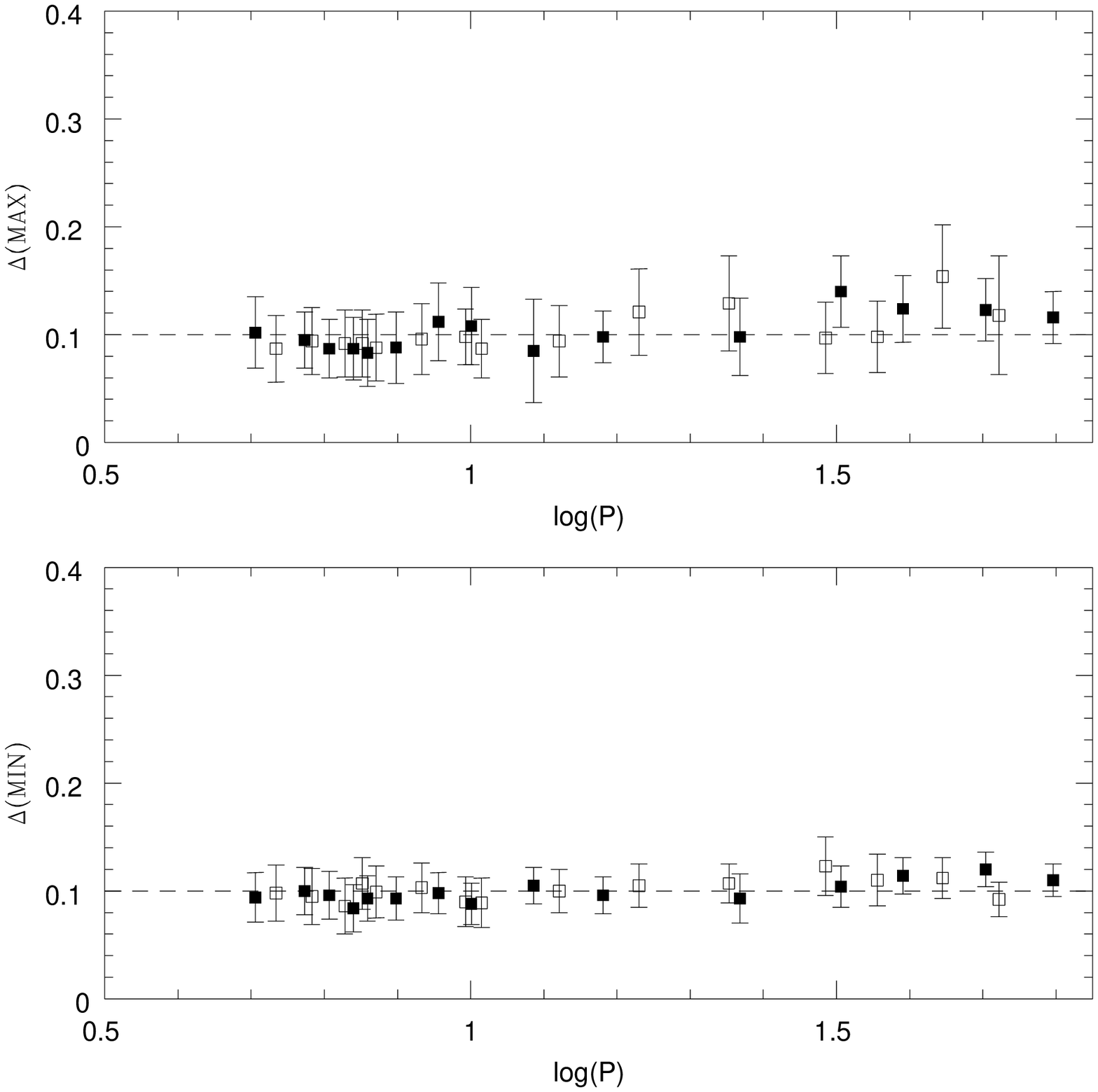}
         \epsfxsize=7.5cm \epsfbox{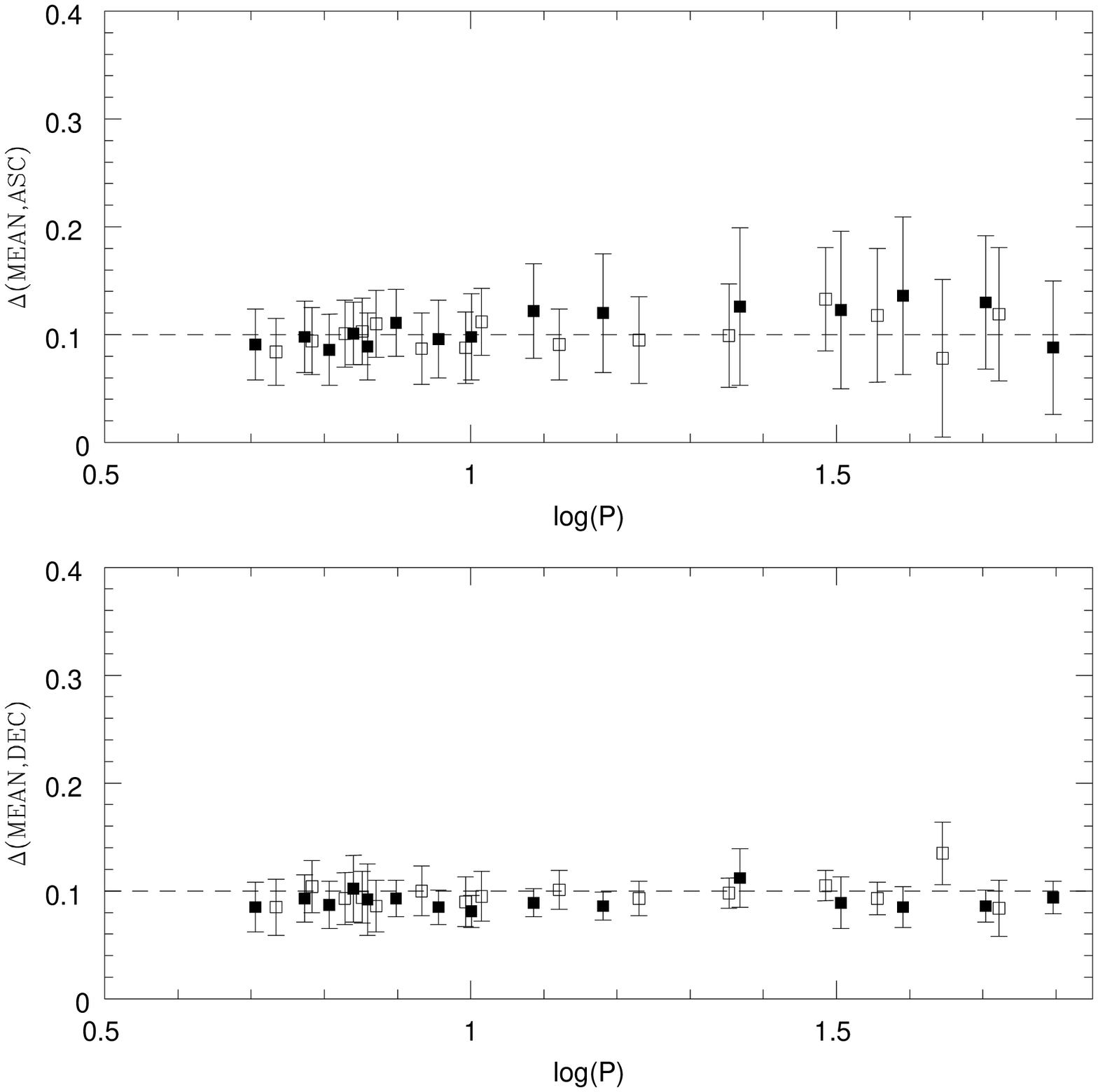}}
       \vspace{0cm}
       \caption{The plots of $\Delta$ as function of $\log(P)$. The open and squares and open circles are the models calculated with \citet{bon00} and \citet{chi89} ML relation, respectively. The dashed lines represent (roughly) the outer boundary of the HIF.}
       \label{delta}
     \end{figure*}

The results from the SMC models were compared to the observations, also presented in Figure \ref{pt}-\ref{ac}. In these figures, Fourier fits to
the SMC Cepheid data, taken from Paper IV, were used to calculate the observed amplitudes and colours at maximum, mean, and minimum light using the definitions given in Paper I \& III. Table \ref{tabpc} \& \ref{tabac} summarizes the empirical PC and AC relations at maximum, mean and minimum light for the SMC Cepheid data plotted in these figures, respectively. In Figure \ref{pt}, we compare the temperatures given in Table \ref{tabmaxmin} \& \ref{tabmean} to the observed SMC Cepheid data. Similarly, Figure \ref{pc} displays the PC relations at various phases for the SMC models superimposed on the SMC data. From these figures, we found that the SMC PC relation at minimum light has the smallest scatter for all periods and does not suffer a statistically
significant change of slope at $\log (P) \sim 1.0$, i.e. it is linear over the entire period range. The PC relation at maximum light is significantly
flatter for long ($\log [P] > 1.0$) period than short period Cepheids and displays marginal evidence of a slope change at
$\log (P) \sim 1.0$. It also has the largest scatter over all phases.

The SMC models do a reasonable job in matching the observations in the period-temperature and the period-colour plane, though the short period models at maximum light tend to be too cool compared to the observations (as in Figure \ref{pt} \& \ref{pc}). Figure \ref{ac} portrays the results of AC relations from the models as compared to the SMC data. Several models, especially those using the \citet{bon00} ML relation, display a smaller amplitude as compared to the data. This is seen in the Galactic (Paper II) and LMC (Paper III) models as well, however we believe this does not affect our results on the HIF-photosphere interactions (Paper III). To the best of our knowledge a comparison of models and theory on PC/AC planes at different phases, as described here, has not been carried out in such detail before: further work will investigate in detail some of the discrepancies between models and observations described in the text.


     \begin{figure*}
       \vspace{0cm}
       \hbox{\hspace{1.2cm}\epsfxsize=7.5cm \epsfbox{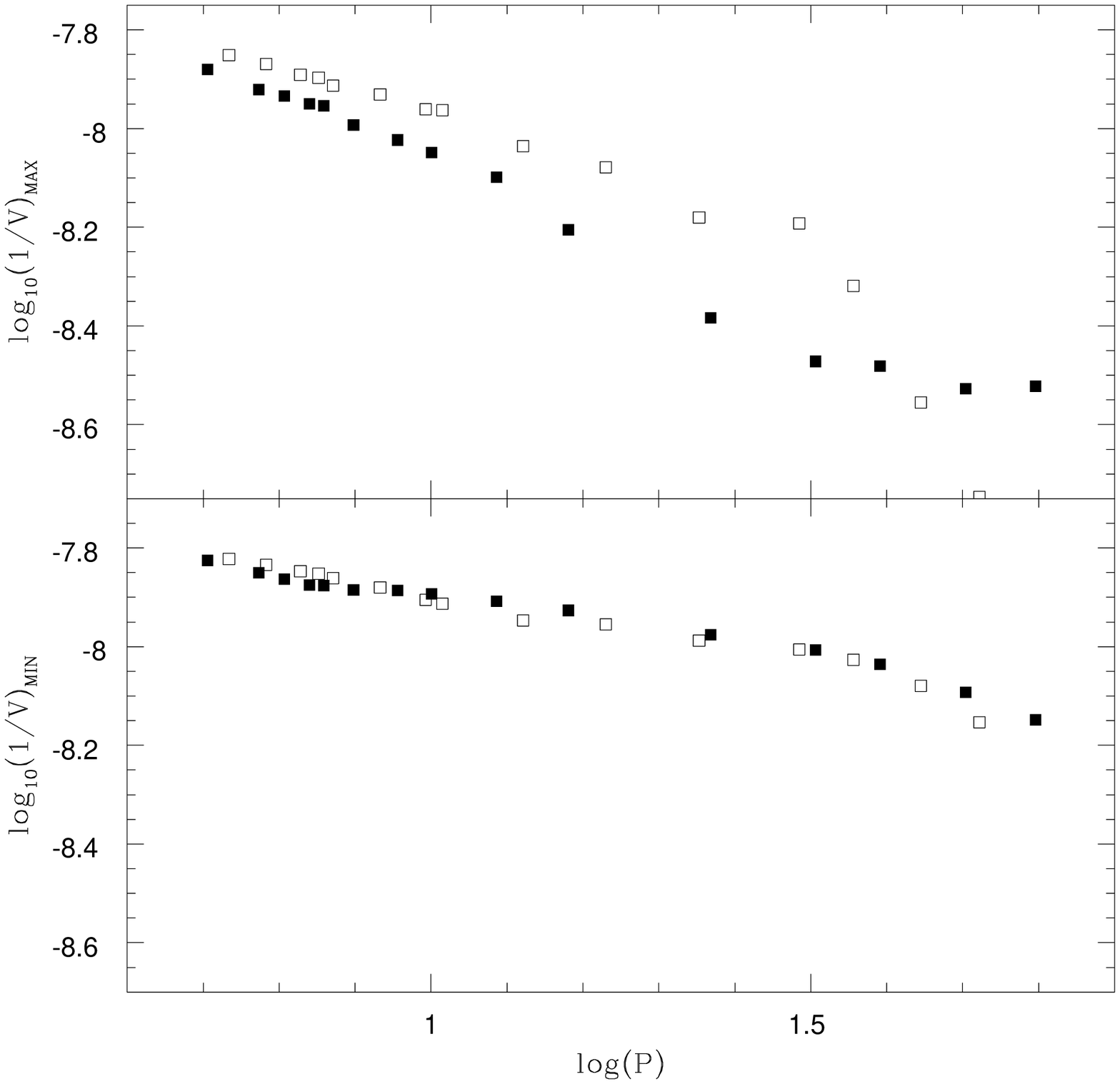}
         \epsfxsize=7.5cm \epsfbox{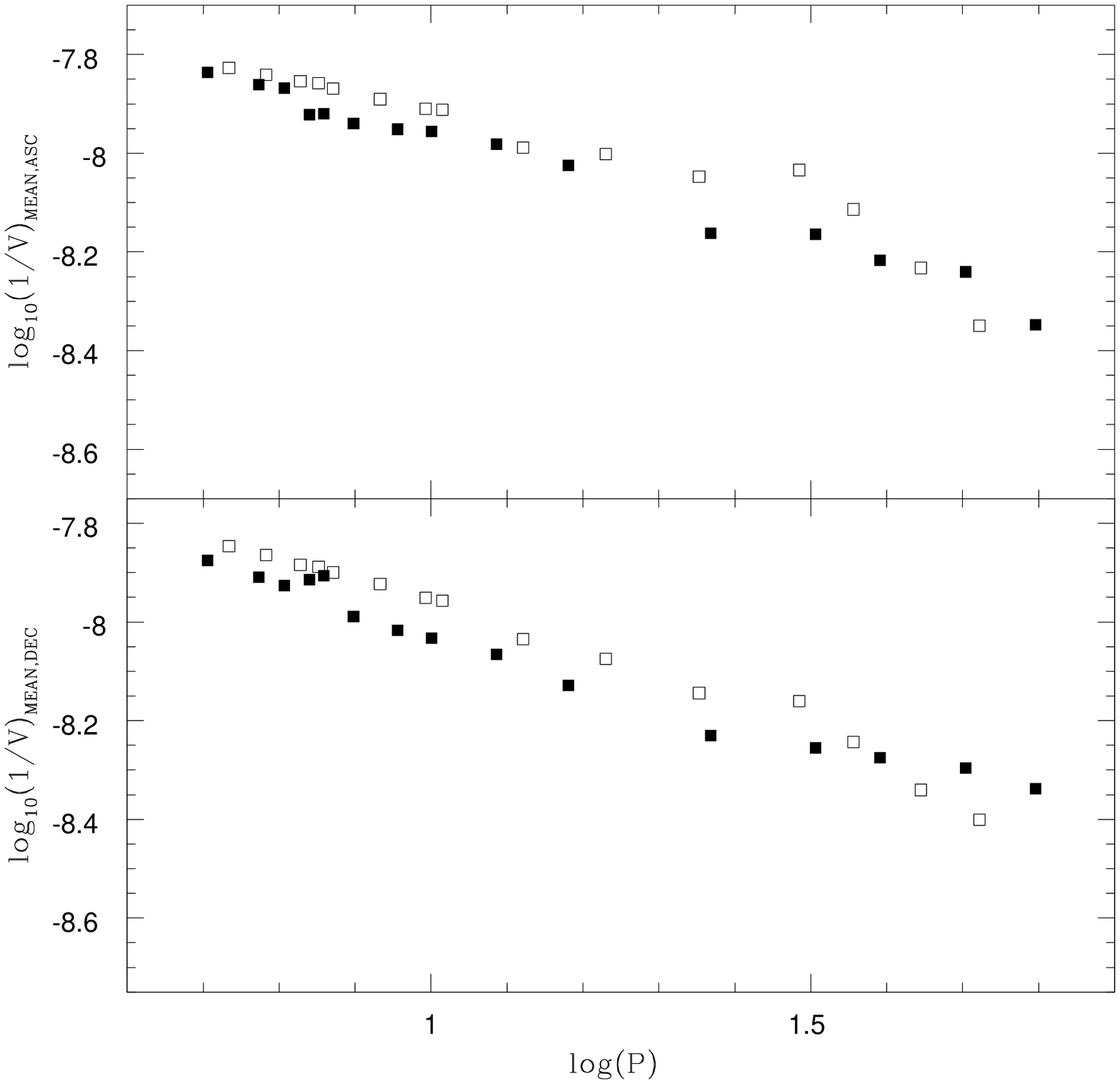}}
       \vspace{0cm}
       \caption{Log of the photospheric density (defined as 1/V where V is the specific volume) plotted against period for the SMC models. The symbols are the same as in Figure \ref{delta}.}
       \label{density}
     \end{figure*}

From the full amplitude SMC models, the temperature and the opacity profile can be plotted at a given phase of pulsation. Similar to Paper II \& III, the locations of the HIF (sharp rise in the temperature profile) and photosphere can be identified
from the temperature profile. Figure \ref{profile} displays the temperature and opacity profiles, with the locations of photosphere marked as filled circles, for a long period, a 10-days and a short period SMC model. From the temperature profile, the ``distance'' $\Delta$  between the HIF and the photosphere from the temperature profile can be calculated (see Paper II for the definition of $\Delta$). A small $\Delta$ implies there is a HIF-photosphere interaction, and vice versa. Figure \ref{delta} presents the $\Delta$ as a function of pulsating period for the SMC models with the two ML relations used in this paper. In Papers II \& III, it is found that the distribution of $\Delta$ as a function of period is almost independent of the adopted ML relations. This is also seen in the SMC models as depicted in Figure \ref{delta}. Furthermore, we see, in contrast to the behavior for Galactic and LMC models, the distance
is reasonably constant for all periods and at all phases. Figure 15 of Paper III shows that at minimum light this distance for 
the Galactic models increases with period, but for the LMC models this distance only increases for periods greater
than 10 days. At maximum light, all models display a constant distance. The SMC models are different in that 
there is almost a constant distance between the HIF and stellar photosphere, independent of pulsation period, pulsation phase and ML
relation.

Figure \ref{density} depicts the photospheric density as a function of period for the SMC models.
This figure needs to be compared with figure 16 of Paper III.
The difference between the SMC models and the Galactic/LMC models is in the photospheric density at minimum light.
The Galactic models have a photospheric density at minimum light which is less than $10^{-8}\mathrm{gcm}^{-3}$. Paper II also found that
for these Galactic models, the photosphere and HIF are not engaged. For the LMC models at minimum light, Paper III found that the
HIF and photosphere are nearly always engaged with some indication that there is a separation for periods greater than 10 days.
We also see that the photospheric density for these longer period LMC models is less than $10^{-8}\mathrm{gcm}^{-3}$.
The SMC models always have an engaged photosphere and HIF at all phases. But crucially, the density of
the photosphere is higher than $10^{-8}\mathrm{gcm}^{-3}$ for models with $\log(P) < 1.5$.
The higher densities mean that no sharp change in the temperature of the photosphere independent of the global stellar
parameters can occur because our premise is that a sharp change can occur in the slope of the PC relation when the photosphere and HIF
are engaged and this occurs at low densities. Thus the PC relation for SMC Cepheids should show a much reduced change in the PC slope in going from short to long period Cepheids and this is what we see in the observations: at maximum light, Galactic and LMC Cepheids display a PC(max) relation with a slope that decreases to zero in going from short to long period (Paper I \& III) but SMC Cepheids only show a reduced PC slope at maximum light. It may be possible to test this assertion spectroscopically and this will be the subject of a future investigation.
Because the PC relation at mean light is
an average of the PC relation at different phases, if there is a discontinuity or not at a certain phase, this effect will be transmitted to
the mean light PC relation and hence to the mean light PL relation.

\section{Conclusion and Discussion}

In this paper we construct the full amplitude models appropriate for the SMC Cepheids to study the PC/AC relations and the HIF-photosphere interaction.
In short, the SMC models constructed in this paper do a reasonable job of matching the observed PC relations at minimum and mean light. The greatest discrepancies occur for short period Cepheids at maximum light. These SMC models tend to not be driven to higher temperatures and hence bluer $(V-I)$ colours, though they do
fall inside the outer envelope of the observed data points.

       \begin{table*}
         \centering
         \caption{Summary of various properties at the maximum and minimum light for the Galactic, LMC and SMC Cepheids and models. Amp and T refer to the amplitude and temperature, respectively.}
         \label{tabsum}
         \begin{tabular}{lllll} \hline
           Galaxy   & Photospheric Density & HIF-Photosphere Interaction & Observed PC Relation & Observed AC Relation \\
           \hline   
           \multicolumn{5}{c}{At Maximum Light} \\
           GAL & Low for $\log(P) \gtrsim 0.7$   & Yes for all periods          & Flat for $\log(P) \gtrsim 0.8$ with  & Higher Amp. $\rightarrow$ hotter T. \&  \\
               &                                 &                              & significant break at $\log(P)=1.0$
  & gets shallower for $\log(P)>1.0$ \\
           \multicolumn{5}{c}{} \\
           LMC & Low for $\log(P) \gtrsim 1.0$   & Yes for all periods          & Flat for $\log(P)> 1.0$ with
  & Higher Amp. $\rightarrow$ hotter T. \&  \\
               & High for $\log(P) \lesssim 1.0$ &                              & significant break at $\log(P)=1.0$
  & gets shallower for $\log(P)>1.0$ \\
           \multicolumn{5}{c}{} \\
           SMC & High for $\log(P) \gtrsim 1.0$   & Yes for all periods          & $\sim$Flatter for $\log(P)> 1.0$ with
  & Higher Amp. $\rightarrow$ hotter T. \&  \\
               & High for $\log(P) \lesssim 1.0$ &                              & marginal break at $\log(P)=1.0$
  & gets shallower for $\log(P)>1.0$ \\

           \multicolumn{5}{c}{} \\
           \multicolumn{5}{c}{At Minimum Light} \\
           GAL & Low for $\log(P) \gtrsim 0.7$   & No for $\log(P) \gtrsim 0.7$   & Non-flat slope for all periods & Higher Amp. $\rightarrow$ cooler T. \&  \\
               &                                 &                                & \& no significant break     & gets shallower for $\log(P)>1.0$ \\
           \multicolumn{5}{c}{} \\
           LMC & Low for $\log(P) \gtrsim 1.0$   & No for $\log(P) \gtrsim 1.0$   & Non-flat slope for all periods,        & Flat for $\log(P)<1.0$ \& \\
               & High for $\log(P) \lesssim 1.0$ & Yes for $\log(P) \lesssim 1.0$ & significant break at $\log(P)=1.0$ & gets steeper for $\log(P)>1.0$  \\
           \multicolumn{5}{c}{} \\
           SMC & Low for $\log(P) \gtrsim 1.5$   & Yes for all periods            & Non-flat slope for all periods         & $\sim$Flat for all periods    \\
               & High for $\log(P) \lesssim 1.5$ &                                & \& no significant break             &                               \\

           \hline
         \end{tabular}
       \end{table*}

From the previous papers in the series we note that the empirical LMC PC relation at minimum light has a statistically significant slope change or ``break'' at $\log (P) \sim 1.0$, but has the smallest scatter over all phases. The LMC PC relation at maximum light has the largest scatter over all phases. It also shows a statistically significant change of slope across $\log (P) \sim 1.0$ but has a flat slope (i.e. close to zero) when only the long period Cepheids are considered. The empirical Galactic PC relation displays a break at 10 days with a flat relation thereafter at the maximum light and a linear relation at minimum light. The empirical SMC PC relation also display a marginal break at maximum light but is linear at minimum light. The PC relation is flatter for the SMC Cepheids with $\log(P)>1$ than the short period SMC Cepheids at the maximum light.

In terms of the empirical AC relation, the AC relations at maximum/minimum light are significantly broken (at $\log[P]=1.0$) for all three galaxies (see Paper I \& III). At maximum light, higher amplitude Cepheids in all three galaxies are driven to {\it hotter} temperature and hence bluer colours for all period ranges (with negative slope in the AC relation). Nevertheless the slope of these empirical AC relations at maximum light becomes shallower for the long period Cepheids. However, the opposite behavior is found at the minimum light as the long period Cepheids are driven to {\it cooler} temperature as the amplitude increases in all three galaxies. For short period Cepheids, the same behavior is found for the Galactic Cepheids but the LMC and SMC short period AC relations are nearly flat at minimum light. In all cases the behavior at mean light is intermediate between the properties at maximum/minimum light. 

The Saha ionization equation, which is used in most models of Cepheid envelopes, is such that when the photosphere and HIF interact at low densities, the photospheric temperature will be almost independent of pulsation period. This is because when the HIF and photosphere are interacting or engaged, the photospheric temperature is essentially the same as the temperature at which hydrogen ionizes, and not dependent on the pulsation period. Hence the colour, either in $(V-I)$ or in $(B-V)$, is also almost independent of the pulsation period. Moreover, as the period increases, the $L/M$ ratio increases which, in turn, changes the location of the HIF, in terms of the mass distribution, in the Cepheid envelope. This can change the phase at which the HIF and photosphere interact. 

One possible difference between Galactic, LMC and SMC models is the photospheric density at minimum light. For Galactic models, the $L/M$ ratio causes the low density photosphere-HIF interaction to only occur at/around the maximum light, leading to a stronger relation between period and colour at minimum light. At other phases, the HIF and photosphere are well separated. For LMC models, the $L/M$ ratio forces the HIF-photosphere interaction to occur at all phases for short period Cepheids at higher densities and at maximum light for longer period Cepheids at low densities. At phases close to the minimum light, the HIF-photosphere become disengaged for longer period Cepheids. This leads to change in the slope of PC relation at phases around minimum light for the long and short period LMC Cepheids. For SMC models, the HIF-photosphere interaction occurs throughout the pulsation cycle, but only occurs at low density at phases close to, and including, the maximum light for the long period models. This leads longer period SMC Cepheids to have a flatter PC relation slope than the short period SMC Cepheids at the maximum light. The high density photosphere-HIF interaction occurs at other phases and ensures that there is no significant change of slope in the PC relation as a function of period. The low density HIF-photosphere interaction at maximum light is the reason for the observed flat PC relation at maximum light for the Galactic and LMC Cepheids and the flatter PC(max) relation for longer period SMC Cepheids.
In Table \ref{tabsum}, we summarize some of the properties discovered in this series of papers for the Galactic, LMC and SMC Cepheids and models. 

Because these changes depend on the $L/M$ ratio and because the $ML$ relation is dependent on metallicity, the above argument
outlines a mechanism whereby the observed changes in the PC relation between Galactic, LMC and SMC Cepheids could occur. Moreover
these changes do not have to be monotonic with metallicity. Since the mean light PC relation is the average of the PC relation at different phases, changes in the slope of the PC relation, can, in principle, affect the mean light PC relation (see Paper I to IV). 

In addition to our interest in understanding the physics behind the multi-phase PC and PL relations as a function of metallicity (in Galaxy, LMC and SMC), we are also interested in the changes of light curves shape in different galaxies. This is because the AC relations are essentially equivalent to plotting the light curve shape against temperature (or colour). Previous work has established that Principal Component Analysis (PCA) is a very efficient way of describing light curve structure with upwards of $80\%$ of structure variation being explained by just the first Principal Component \citep{kan02,tan05}. Since the first PCA coefficient correlates well with amplitude, therefore there will be a correlation between the amplitude and light curve structure. We remark in passing that \citet{kel06} 
mention a correlation between amplitude and effective temperature in their hydrodynamic models of Cepheids. It can be seen from this work and
previous papers in this series that a correlation between amplitude and effective temperature is due to a more extreme correlation at maximum or minimum light:
the physics of which has been explained in this paper and earlier papers in the series.


\section*{acknowledgments}

We thank an anonymous referee for useful suggestions. We also thank T. Barnes for helpful discussion. SMK acknowledges the support from HST-AR-10673.04-A and Chretien International Research Award. CN acknowledges support from NSF award OPP-0130612 and a University of Illinois seed funding award to the Dark Energy Survey.
        

\end{document}